\documentstyle[12pt]{article}
\renewcommand{\theequation}{\thesection.\arabic{equation}}

\newcommand{\NP}[1]{Nucl. \ Phys.}
\newcommand{\PL}[1]{Phys. \ Lett.}
\newcommand{\p}[1]{\partial}
\newcommand{\PRL}[1]{Phys.\ Rev.\ Lett. }
\newcommand{\MPL}[1] { Mod. Phys. Lett. }
\newcommand{\IJMP}[1] { Int. J. Mod. Phys. }
\begin{document}

\title{p-Brane Solutions in Diverse Dimensions}
\author{
I.Ya.Aref'eva \thanks{Permanent address: Steklov Mathematical Institute,
Vavilov St.42, GSP-1, 117966, Moscow,
e-mail: arefeva@class.mian.su},$~~$ K.S.Viswanathan \thanks{e-mail:
kviswana@sfu.ca} \\$~$
 and $~$  \\I.V.Volovich \thanks {Permanent address: Steklov Mathematical
Institute,
Vavilov St.42, GSP-1, 117966, Moscow, e-mail: volovich@class.mian.su}\\
$~$\\ Department of Physics, Simon Fraser University\\
Burnaby, British Columbia, V5A 1S6, Canada}
\date {$~$}
\renewcommand{\theequation}{\thesection.\arabic{equation}}
\maketitle
\begin{abstract}
A generic Lagrangian, in arbitrary spacetime
dimension, describing the interaction of a graviton,
a dilaton and two antisymmetric tensors is considered. 
An isotropic p-brane solution consisting of three
blocks and depending on four parameters in the Lagrangian
and two arbitrary harmonic functions
is obtained. For specific values of parameters
in the Lagrangian the solution may be identified with 
previously known superstring solutions.
\end{abstract}
\newpage
 \section{Introduction} 
Recent remarkable developments in superstring theory
lead to the discovery that five known superstring
theories in ten dimensions are related by duality
transformations  and that there are also $M$-theory
in $11d$ and $F$-theory in $12d$ that are useful
in the study of the moduli space of quantum vacua \cite {FILQ}
-\cite{Vaf}.
Duality requires  the presence of extremal black holes
in the superstring spectra. A derivation of the Bekenstein-Hawking
formula for the entropy of certain extremal black holes was given
using the D-brane approach \cite{SV}-\cite{pol2}. 
In all these developments
the study of p-brane solutions of the supergravity equations
play an important role \cite{DGHR}-\cite{berg}.

In this paper we shall consider the following action
\begin{equation}
 I=\frac{1}{2\kappa ^{2}}\int d^{D}x\sqrt{-g}
 (R-\frac{1}{2}(\nabla  \phi) ^{2}
 -\frac{1}{2(q+1)!}e^{-\alpha \phi}F^{2}_{q+1}
 -\frac{1}{2(d+1)!}e^{\beta \phi}G^{2}_{d+1}),
      \label{011}
 \end{equation}
It describes the interaction of the gravitation field $g_{MN}$
with the dilaton $\phi$ and with two antisymmetric fields:
$F_{q+1} $ is a closed $q+1$-differential form and  $G_{d+1}$
is a closed $d+1$-differential form. Various supergravity
theories contain the terms from  (\ref{011}).

The aim of this paper is to present a solution of (\ref{011})
with the metric of the form
\begin{equation}
ds^{2}=H_{1}^{-2a_{1}}H_{2}^{-2a_{2}}\eta_{\mu \nu} dy^{\mu} dy^{\nu}+
H_{1}^{-2b_{1}}H_{2}^{-2a_{2}}\delta_{nm}dz^{n}dz^{m}+
H_{1}^{-2b_{1}}H_{2}^{-2b_{2}}\delta_{\alpha\beta}dx^{\alpha}dx^{\beta}, 
 \label{12}
\end{equation}
where $\eta_{\mu\nu}$ is a flat Minkowski metric,
$$\mu, ~\nu = 0,...,q-1;~~m,n=1,2,...,d-q,$$ and 
$$\alpha,\beta =1,...,D-d.$$
  For definitness we assume  that $ D>d\geq q$.

The parameters $a_{i}$ and
 $b_{i}$ in the solution (\ref{12}) are rational functions of 
the parameters in the action (\ref{011}):
\begin{equation}
a_{1}=\frac{2\tilde q}{\alpha^{2}(D-2)+2q\tilde q},
~~~~a_{2}=\frac{\alpha^{2}(D-2)}
{\alpha^{2}d(D-2)+2\tilde d q^{2}}
 \label{13}
\end{equation}
\begin{equation}
b_{1}=-\frac{2q}{\alpha^{2}(D-2)+2q\tilde q},
~~~~b_{2}=-\frac{\alpha^{2}d(D-2)}{\tilde d[
\alpha^{2}d(D-2)+2\tilde d q^{2}]}
\label{14}
\end{equation}
where
\begin{equation}
\tilde d=D-d-2,~~~~\tilde q=D-q-2.
\label{14a}
\end{equation}

Our solution  (\ref{12}) is valid only if the following
relation between parameters in the action is satisfied
\begin{equation}
\alpha\beta=\frac{2q\tilde d}{D-2}
\label{15}
\end{equation}
There are   two arbitrary harmonic functions
  $H_{1}$ and $H_{2}$ of variables $x^{\alpha}$ in (\ref{12}),
 \begin{equation}
  \Delta H_{1}=0,~~~~~\Delta H_{2}=0.
  \label{16}
 \end{equation}
 
Non-vanishing components of the differential form are given by
 \begin{equation}
  {\cal A}_{\mu_{1}...\mu_{q}}=h
  {\epsilon_{\mu_{1}...\mu_{q}}}H_{1}^{-1},~~~~ F=d{\cal A},
  \label{18}
 \end{equation}
 \begin{equation}
  {\cal B}_{I_{1}...I_{d}}=k
  \epsilon_{I_{1}...I_{d}}H_{2}^{-1}, ~~~G=d{\cal B},~~~I=0,...d-1.
  \label{18a}
 \end{equation}
Here $\epsilon _{123..,q}=1$, $\epsilon _{123...d}=1$ and
$h$ and $k$ are given by the formulae 
\begin{equation}
h^{2}=\frac{4(D-2)}
{\alpha ^{2}(D-2)+2q\tilde q},
 \label{194}
\end{equation}
\begin{equation}
k^{2}=\frac{2\alpha^{2} (D-2)^{2}}
{{\tilde d}[\alpha ^{2}d(D-2)+2q^{2}{\tilde d}]},
 \label{195n}
\end{equation}

The dilaton field is
\begin{equation}
  \phi=\frac{1}{2}\beta k^{2}\ln H_{2}-\frac{1}{2}
  \alpha h^{2}\ln  H_{1}.
  \label{17}
\end{equation}
We obtain the solution  (\ref{12}) by reducing the Einstein
equations to the system of algebraic equations. To this end
we introduce a linear dependence between functions
in the Ansatz (see below). 
 The solution (\ref{12}) consists of three blocks, the first block 
 consists of variables $y$, another of variables $z$
 and the other of variables $x$ and all functions depend only on $x$. 
We shall call it
the three-block p-brane solution. 
 
We shall consider also the following  "dual" action
\begin{equation}
\tilde I=\frac{1}{2\kappa ^{2}}\int d^{D}x\sqrt{-g}
 (R-\frac{1}{2}(\nabla  \phi) ^{2}
 -\frac{1}{2(q+1)!}e^{-\alpha \phi}F^{2}_{q+1}
 -\frac{1}{2({\tilde d}+1)!}e^{\tilde\beta \phi}G^{2}_{{\tilde d}+1}),
      \label{11}
 \end{equation}
where  $G_{\tilde d+1}$ is a closed $\tilde d+1$-differential form.
If $\tilde{d}$ is related to $d$ by (\ref {14a})  and 
\begin{equation}
\tilde\beta =-\beta
      \label{11a}
 \end{equation}
then the  solution for the metric (\ref{12}) with the differential
form $F$ (\ref{18}) and the dilaton (\ref{17}) is 
valid also  for the action (\ref{11}). An expression for 
the antisymmetric
field $G$ will be different, namely
 \begin{equation}
   G^{\alpha_{1}...\alpha _{\tilde d+1}}=k
   H_{1}^{\sigma_{1}}H_{2}^{\sigma_{2}}
   \epsilon ^{\alpha _{1}...\alpha_{\tilde d+1}\beta}
   \partial _{\beta} H_{2}^{-1}. 
  \label{1999}
 \end{equation}
here $\epsilon ^{123..\tilde d+2}=1$ and
\begin{equation}
\sigma_{1}=\frac{\alpha\beta h^{2}}{2}(1-\frac{1}{\tilde d }),
~~~~\sigma_{2}=\frac{\beta k^{2}}{2}(\frac{1}{\tilde d }-1)
\label{917}
\end{equation}

    The three-block p-branes solution
for the Lagrangian with one differential form for various dimensions
of the space-time  was found in \cite{AV}. It 
contains previously known D=10 case \cite{Tseytlin,CM}.
Equations of motion for the case of one form corresponds to equation of 
motion for ansatz (\ref{12}), (\ref{18}) and (\ref{1999}) for the 
dual action (\ref{11}) when $\alpha$=
$\beta$ and $q=\tilde{d}$.

Note that the metric (\ref{12}) describes also the solution for the 
action with 
the form $F_{q+1}$ replaced by its dual $F_{\tilde{q}+1}$ with
$\tilde{q}+q+2=D$  and $\alpha \to \tilde{\alpha}=-\alpha$.
One can also change two forms $F$ and $G$ to their dual version
without changing the metric (\ref{12}).

To illustrate our method
on a simple example we first consider in Sect. 3  the simple case when
in (\ref{011})
$d =q$ and one has only two blocks in the metric.
Then in Sect. 4 we derive the solution (\ref{12}). In Sect. 5
we consider particular cases of the solution (\ref{12})
and obtain different known solutions. In Appendix  the solution
of the system
of algebraic equations is given.

\section{Two block solution}
To illustrate the method of solution in this section
we consider the simple case when the system of
algebraic equations can be easily solved. Let us consider the action
\begin{equation}
I=\frac{1}{2\kappa ^{2}}\int d^{D}z\sqrt{-g}
[R-\frac{1}{2}(\partial \phi) ^{2}-
 \frac{e^{-\alpha \phi}}{2(d+1)!}F^{2}_{d+1} 
-\frac{e^{\beta \phi}}{2(d+1)!}G^{2}_{d+1}) 
  \label{31}
\end{equation} 
      The Einstein equations  for the action (\ref{11}) read
\begin{equation}
  R_{MN}-\frac{1}{2}g_{MN}R=T_{MN},
  \label{19a}
\end{equation}
where the energy-momentum tensor has the form   
$$T_{MN}=  \frac{1}{2}
  (\partial _{M}\phi \partial _{N}\phi -
  \frac{1}{2}g_{MN} (\partial \phi)^{2}) 
$$
$$
    + \frac{1}{2d!}e^{-\alpha \phi}
( F_{MM_{1}...M_{d}}F_{N}^{M_{1}...M_{d}}-
 \frac{1}{2(d+1)}g_{MN}F^{2})+
 $$

\begin{equation}
  \frac{1}{2d!}e^{\beta \phi}
(G_{MM_{1}...M_{d}}G_{N}^{M_{1}...M_{d}}-
 \frac{1}{2(d+1)}g_{MN}G^{2}) 
 \label{32}
\end{equation} 
and one has also equations of motion

\begin{equation}
\partial _{M}(\sqrt{-g}e^{-\alpha \phi}F^{MM_{1}...M_{d}})=0 
 \label{33}
\end{equation} 

\begin{equation}
\partial _{M}(\sqrt{-g}e^{\beta \phi}G^{MM_{1}...M_{d}})=0 
 \label{34}
\end{equation} 

\begin{equation}
\partial _{M}(\sqrt{-g}g^{MN}\partial _{N }\phi ) 
+ 
\frac{\alpha }{2(d+1)!}\sqrt{-g}e^{-\alpha \phi}F^{2}-
\frac{\beta }{2(d+1)!}\sqrt{-g}e^{\beta \phi}G^{2}=0
 \label{35}
\end{equation} 

We use the following Ansatz
\begin{equation}
F=d{\cal A} , ~~~~~{\cal A}_{01...d-1}=\gamma _{1}e^{C_{1}(x)}
 \label{36}
\end{equation} 

\begin{equation}
G=d{\cal B} ,~~~~~ {\cal B}_{01...d-1}=\gamma _{2}e^{C_{2}(x)}
  \label{37}
\end{equation} 

\begin{equation}
ds^{2}=e^{2A(x)}\eta_{\alpha \beta} dy^{\alpha} dy^{\beta}
+e^{2B(x)}dx^{i}dx^{i}, 
 \label{38}
\end{equation} 

$\alpha$, $\beta$ =0,1,...,d-1, $\eta_{\alpha \beta}$
is a flat Minkowski metric, $i,j$ =d,...,D.

With the above Ansatz equations (\ref{19a}) are reduced  to
the following system of equations:
\begin{eqnarray}
 (d-1)\Delta A+
 (\tilde{d}+1)\Delta B +
 \frac{d(d-1)}{2}(\partial A)^{2} +
 \frac {\tilde{d}(\tilde{d}+1)}{2}(\partial B)^{2}+
(d-1)\tilde{d}(\partial A\partial B) = \nonumber \\
 -\frac{\gamma _{1}^{2}}{4}e^{\alpha \phi -2dA +2c_{1}}(\partial 
 c_{1})^{2}-
   \frac{\gamma _{2}^{2}}{4}e^{\alpha _{2}\phi -2dA +2c_{2}}(\partial 
       c_{1})^{2}-
      \frac{1}{4}(\partial \phi)^{2} 
      \label{39}
\end{eqnarray}
and
$$
-d(\partial _{m}\partial _{n} A +\partial_{m}A \partial _{n}A )
-{\tilde d}(\partial _{m}\partial _{n}B
-\partial _{m}B\partial _{n}B)
+d(\partial _{m}A\partial _{n} B+\partial _{m}B\partial _{n} A)$$

$$
+\delta _{mn}[d\Delta A + \frac{d(d+1)}{2}(\partial A)^{2}
+{\tilde d}\Delta B+ \frac{{\tilde d}({\tilde d}+1)}{2}(\partial B)^{2}
+d({\tilde d}-1)(\partial A \partial B)]=$$

$$\frac{1}{2}[\partial _{m}\phi\partial_{n} \phi -\frac{1}{2}\delta_{mn}
(\partial \phi)^{2}]
-\frac{\gamma _{1}^{2}}{4}e^{\alpha \phi -2dA +2C_{1}}
[\partial _{m}C_{1}\partial _{n}C_{1} -\frac{1}{2}\delta _{mn}(\partial 
C_{1})^{2}]
$$

\begin{equation}
-\frac{\gamma _{2}^{2}}{4}e^{-\beta \phi -2dA +2C_{2}}
[-\partial _{m}C_{2}\partial _{n}C_{2} -\frac{1}{2}\delta _{mn}(\partial 
C_{2})^{2}]
  \label{310}
\end{equation}
Here  $\tilde d=D-d-2$.
Equations for antisymmetric fields are

\begin{equation}
\partial _{m}(e^{-\alpha  \phi -dA +{\tilde d}B +c_{1}}\partial 
_{m}c_{1})=0
 \label{311}
\end{equation} 

\begin{equation}
\partial _{m}(e^{\beta  \phi -dA +{\tilde d}B +c_{1}}\partial 
_{m}c_{2})=0
  \label{312}
\end{equation} 

Equation of motion for dilaton 

$$\partial _{m}(e^{-dA +{\tilde d}B}\partial _{m}\phi )
-\frac{\alpha\gamma _{1}^{2}}{2}e^{-\alpha \phi -2dA +2C_{1}}
(\partial _{m}C_{1})^{2}$$

\begin{equation}
 +\frac{\beta \gamma _{1}^{2}}{2}e^{\beta \phi -2dA +2C_{2}}
(\partial _{m}C_{2})^{2} =0
  \label{31m}
\end{equation} 

We impose the following relations:
\begin{equation}
dA +{\tilde d}B =0,
 \label{314}
\end{equation} 

\begin{equation}
-\alpha \phi -2dA +2C_{1}=0,
 \label{315}
\end{equation} 

\begin{equation}
\beta \phi -2dA +2C_{2}=0.
 \label{316}
\end{equation} 

Under these conditions equations  (\ref{311}), (\ref{312}) and (\ref{31m})
take the following forms, respectively,

\begin{equation}
\partial _{m}(e^{-C_{1}}\partial _{m}C_{1})=0,
                  \label{317}
\end{equation} 

\begin{equation}
\partial _{m}(e^{-C_{2}}\partial _{m}C_{2})=0,
                  \label{318}
\end{equation} 

\begin{equation}
\Delta \phi 
-\frac{\alpha \gamma _{1}^{2}}{2}(\partial _{m}  C _{1})^{2}
+\frac{\beta \gamma _{2}^{2}}{2}(\partial _{m}  C _{2})^{2}=0
                  \label{319}
\end{equation} 

>From equations (\ref{317}) , (\ref{318})  and (\ref{319}) we get
\begin{equation}
 \Delta C _{1} 
=(\partial C_{1})^{2},
                  \label{320}
\end{equation} 
\begin{equation}
 \Delta C _{2} 
=(\partial C_{2})^{2}
                  \label{321}
\end{equation} 
and
\begin{equation}
\phi=\varphi _{1}C_{1}+\varphi _{2}C_{2}
                  \label{322}
\end{equation} 
with 
\begin{equation}
\varphi _{1}=\frac{\alpha \gamma ^{2}}{2}
                  \label{323}
\end{equation} 
\begin{equation}
\varphi _{2}=-\frac{\beta\gamma ^{2}}{2}
                  \label{324}
\end{equation} 

Equations (\ref{315}) and (\ref{316}) give 
\begin{equation}
A=a_{1}C_{1}+a_{2}C_{2}, 
                    \label{325}
\end{equation}

where 
\begin{equation}
a_{1}=\frac{\beta }{d(\alpha +\beta )},~~~
a_{2}=\frac{\alpha }{d(\alpha +\beta )}, 
                    \label{326}
\end{equation} 
and 
\begin{equation}
\phi=\varphi _{1}C_{1}+\varphi _{2}C_{2}
                  \label{532}
\end{equation} 
where
\begin{equation}
\varphi _{1}=\frac{2}{\alpha +\beta },~~~
                 \label{327}
\end{equation}
\begin{equation}
\varphi _{2}=-\frac{2}{\alpha +\beta }.
                 \label{328}
\end{equation}

Comparing (\ref{323}) with (\ref{327}) and (\ref{324}) with (\ref{328})
we conclude that 
\begin{equation}
 \gamma _{1}^{2}=\frac{4}{\alpha (\alpha +\beta )},~~~
 \gamma _{2}^{2}=\frac{4}{\beta (\alpha +\beta )}.
 \label{329}
\end{equation}

Let us now consider equation (\ref{310}). Since $C_{1}$ and $C_{2}$
are two independent functions we have to have that the terms with
$\partial _{m}C_{1}\partial _{n}C_{2}$ vanish, i.e. we have to 
impose  the  condition
\begin{equation}
 \alpha \beta=\frac{2d{\tilde d}}{d+{\tilde d}}.
  \label{330}
\end{equation} 
  
 Therefore, the Ansatz  (\ref{37}) is consistent with the 
  metric (\ref{38}) only under condition (\ref{330}).  Straitforward 
 calculations show that for $A$,   and $\phi$  given by 
 (\ref{325})-(\ref{326}) and (\ref{532}) in terms of two independent 
 functions $C_{1}$ and $C_{2}$, satisfying equations 
 (\ref{320}) and  (\ref{321}), equations (\ref{39}) (\ref{310})
 are satisfied if we impose the condition (\ref{330}).
 
 Let us write the metric in terms of two harmonic functions:
 \begin{equation}
 H_{1}=-\ln C_{1},~~~  H_{2}=-\ln C_{2} 
  \label{331}
 \end{equation} 
 We finally get 
$$ds^{2}=H_{1}^{-\frac{2\beta }{d(\alpha +\beta )}}
H_{2}^{-\frac{2\alpha }{d(\alpha +\beta )}}
\eta_{\alpha \beta} dy^{\alpha} dy^{\beta}
$$

 \begin{equation}
+H_{1}^{\frac{2\beta }{{\tilde d}(\alpha +\beta )}}
H_{2}^{\frac{2\alpha }{{\tilde d}
(\alpha +\beta )}}dx^{i}dx^{i} 
 \label{332}
 \end{equation} 



\renewcommand{\theequation}{\thesection.\arabic{equation}}
 
\section{Three block solution} 
 \setcounter{equation}{0}

      In this secction we consider equations of motion
for the action  (\ref{11}).
  The Einstein equations  for the action (\ref{11}) read

\begin{equation}
  R_{MN}-\frac{1}{2}g_{MN}R=T_{MN},
  \label{19}
\end{equation}
where the energy-momentum tensor is
$$   
 T_{MN}= \frac{1}{2}
  (\partial _{M}\phi \partial _{N}\phi -
  \frac{1}{2}g_{MN} (\partial \phi)^{2}) 
$$

$$
 + \frac{1}{2q!}e^{-\alpha \phi}(F_{MM_{1}...M_{q}}F_{N}^{M_{1}...M_{q}}-
 \frac{1}{2(q+1)}g_{MN}F^{2})
$$
\begin{equation}
 + \frac{1}{2{\tilde d}!}e^{-\alpha \phi}(G_{MM_{1}...M_{{\tilde d}}}
  G_{N}^{M_{1}...M_{{\tilde d}}}-
 \frac{1}{2({\tilde d}+1)}g_{MN}G^{2})
  \label{110}
\end{equation}
The equation of motion for the antisymmetric fields are
\begin{equation}
\partial _{M}(\sqrt{-g}e^{-\alpha \phi}F^{MM_{1}...M_{q}})=0, 
                                  \label{111a}
\end{equation}

\begin{equation}
\partial _{M}(\sqrt{-g}e^{-\beta \phi}G^{MM_{1}...M_{{\tilde d}}})=0, 
                                  \label{111}
\end{equation}
and one has the Bianchi identity
\begin{equation}
 \epsilon ^{M_{1}...M_{q+2}}\partial_{M_{1}}F_{M_{2}...M_{q+2}}=0.
 \label{112}
\end{equation}
 
 \begin{equation}
 \epsilon ^{M_{1}...M_{{\tilde d}+2}}
 \partial_{M_{1}}G_{M_{2}...M_{{\tilde d}+2}}=0.
 \label{1129}
\end{equation}

The equation of motion for the dilaton is
\begin{equation}
\partial _{M}(\sqrt{-g}g^{MN}\partial _{N }\phi ) 
+ 
\frac{\alpha}{2(q+1)!}\sqrt{-g}e^{-\alpha \phi}F^{2}
+ 
\frac{\beta }{2({\tilde d}+1)!}\sqrt{-g}e^{-\beta \phi}G^{2}
=0.
 \label{113}
\end{equation}

We shall solve equations (\ref{19})-(\ref{113})  by using 
the following Ansatz for the metric 

\begin{equation}
ds^{2}=e^{2A(x)}\eta_{\mu \nu} dy^{\mu} dy^{\nu}+
e^{2F(x)}\delta_{nm}dz^{n}dz^{m}+
e^{2B(x)}\delta_{\alpha\beta}dx^{\alpha}dx^{\beta}, 
 \label{114}
\end{equation}
where $\mu$, $\nu$ = 0,...,q-1, $\eta_{\mu\nu}$
is a flat Minkowski metric, $m,n$=$1,2,...,r$ and 
 $\alpha,\beta$ =$1,...,{\tilde d}+2$. Here $A$, $B$
 and $C$ are functions on $x$; $\delta_{nm}$ and $\delta_{\alpha\beta}$
are Kronecker symbols. 

Non-vanishing components of the  differential forms are

\begin{equation}
{\cal A}_{\mu_{1}...\mu_{q}}=h{\epsilon_{\mu_{1}...\mu_{q}}}e^{C(x)}, 
~F=d{\cal A}
 \label{115}
\end{equation}
\begin{equation}
 G^{\alpha_{1}...\alpha _{{\tilde d}+1}}=\frac{1}{\sqrt{-g}}
 k e^{\beta 
 \phi}\epsilon ^{\alpha_{1}...\alpha_{{\tilde d}+1}\gamma}
 \partial _{\gamma} e^{\chi} , 
 \label{116}
\end{equation}
where $h$ and $k$ are constants.
 The left hand side of the Einstein equations for the metric
(\ref{114}) read

$$R_{\mu\nu}-\frac{1}{2}g_{\mu\nu}R=\eta_{\mu\nu}e^{2(A-B)}[(q-1)\Delta A
+
({\tilde d}+1)\Delta B + r\Delta F  $$

$$+\frac{q(q-1)}{2}(\partial A)^{2}+
\frac{r(r+1)}{2}(\partial F)^{2}+\frac{{\tilde d}({\tilde 
d}+1)}{2}(\partial B)^{2}$$

\begin{equation}
+{\tilde d}(q-1)(\partial A\partial 
B)+r(q-1)(\partial A\partial F) +r{\tilde d}(\partial F\partial B)],
 \label{175}
\end{equation}

$$R_{mn}-\frac{1}{2}g_{mn}R=\delta_{mn}e^{2(F-B)}[q\Delta A+
({\tilde d}+1)\Delta B + (r-1)\Delta F $$

$$ +\frac{q(q+1)}{2}(\partial A)^{2}+
\frac{r(r-1)}{2}(\partial F)^{2}+\frac{{\tilde d}({\tilde 
d}+1)}{2}(\partial B)^{2}$$

\begin{equation}
+{\tilde d}q(\partial A\partial 
B)+q(r-1)(\partial A\partial F) +{\tilde d}(r-1)(\partial F\partial B)],
 \label{176}
\end{equation}

$$R_{\alpha \beta} -\frac{1}{2} g_{\alpha \beta} R 
=-q\partial_{\alpha}\partial_{\beta} A-
{\tilde d}\partial_{\alpha}\partial_{\beta} B-
r\partial_{\alpha}\partial_{\beta} F$$

$$-q\partial_{\alpha} A\partial_{\beta}A+
{\tilde d}\partial_{\alpha} B\partial_{\beta}B-
r\partial_{\alpha} F\partial_{\beta} F$$

$$+q(\partial_{\alpha} A\partial_{\beta} B+
\partial_{\alpha} B\partial_{\beta} A)+
r(\partial_{\alpha} B\partial_{\beta} F+
\partial_{\alpha} F\partial_{\beta} B)$$

$$+\delta_{\alpha\beta}[q\Delta A+
{\tilde d}\Delta B + r\Delta F + \frac{q(q+1)}{2}(\partial A)^{2}+
\frac{r(r+1)}{2}(\partial F)^{2}+
\frac{{\tilde d}({\tilde d}-1)}{2}(\partial B)^{2}$$

 \begin{equation}
q({\tilde d}-1)(\partial A\partial B)+
qr(\partial A\partial F) +r({\tilde d}-1)(\partial F\partial B)].
 \label{177}
\end{equation}
For more details see \cite{AV}. Now one  reduces
the  $(\mu\nu)$-components of  
 (\ref{19}) to the equation 

\begin{eqnarray}
 (q-1)\Delta A+
 ({\tilde d}+1)\Delta B +r\Delta F  \nonumber \\ 
 +\frac{q(q-1)}{2}(\partial A)^{2} + \frac{r(r+1)}{2}(\partial F)^{2}+
 \frac {{\tilde d}({\tilde d}+1)}{2}(\partial B)^{2} \nonumber \\ 
+{\tilde d}(q-1)(\partial A\partial B) + r(q-1)(\partial A\partial F) + 
r{\tilde d}(\partial B\partial F) = \nonumber \\ 
 -\frac{1}{4}(\partial 
 \phi)^{2}
 -\frac{{h}^{2}}{4}(\partial C)^{2}e^{-\alpha\phi-2qA+2C}
  -\frac{k^{2}}{4}(\partial \chi)^{2}
 e^{2{\tilde d}B+\beta\phi+2\chi} , \label{117}
\end{eqnarray}

$(nm)$-components of (\ref{19}) to the following equation:

$$q\Delta A+({\tilde d}+1)\Delta B+(r-1)\Delta F$$
$$+\frac{q(q+1)}{2}(\partial A)^{2}+
\frac{{\tilde d}({\tilde d}+1)}{2}(\partial B)^{2}+
\frac{r(r-1)}{2}(\partial F)^{2}$$

$$+q{\tilde d}(\partial A\partial B)+q(r-1)(\partial A\partial F)+
{\tilde d}(r-1)(\partial 
B\partial F)=$$
\begin{equation}
 -\frac{1}{4}(\partial \phi)^{2}+\frac{h^{2}}{4}(\partial 
 C)^{2}e^{-\alpha\phi-2qA+2C}-
 \frac{k^{2}}{4}(\partial \chi)^{2}e^{2{\tilde d}B+\beta\phi+2\chi},
 \label{118}
\end{equation}

and $(\alpha\beta)$-components to the equation:

$$ -q\partial_{\alpha}\partial_{\beta} A-
  {\tilde d}\partial_{\alpha}\partial_{\beta} B-
r\partial_{\alpha}\partial_{\beta} F $$
$$-q\partial_{\alpha} A\partial_{\beta} A + {\tilde d}\partial_{\alpha} 
B\partial_{\beta} B - r\partial_{\alpha} F\partial_{\beta} F + 
q(\partial_{\alpha} A\partial_{\beta} B +\partial_{\alpha} 
B\partial_{\beta} A)$$
$$+r(\partial_{\alpha} B\partial_{\beta} F + \partial_{\alpha} 
F\partial_{\beta} B) + \delta_{\alpha\beta}[q\Delta A+{\tilde d}\Delta B+
r\Delta F$$
$$+\frac{{\tilde d}({\tilde d}+1)}{2}(\partial A)^{2} +
 \frac{r(r+1)}{2}(\partial F)^{2}  $$
$$ +\frac{{\tilde d}({\tilde d}-1)}{2}(\partial B)^{2} +
q({\tilde d}-1)(\partial A\partial B) + 
r({\tilde d}-1)(\partial F\partial B )+ qr(\partial A\partial F)]  $$

$$ = \frac{1}{2}[\partial_{\alpha} \phi\partial_{\beta} \phi - 
\frac{1}{2}\delta_{\alpha\beta}(\partial\phi)^{2}] - 
\frac{h^{2}}{2}e^{-\alpha\phi-2qA+2C}[\partial_{\alpha} C\partial_{\beta} C
 - 
\frac{\delta_{\alpha\beta}}{2}(\partial C)^{2}]$$

\begin{equation}
-\frac{k^{2}}{2} e^{2{\tilde d}B+\beta\phi+
2\chi}[\partial_{\alpha} \chi\partial_{\beta} \chi 
 -\frac{\delta_{\alpha\beta}}{2}(\partial \chi)^{2}],
 \label{119}
\end{equation}
where we use notations 
$( \partial A\partial B)=\partial_{\alpha} A\partial_{\alpha} B$
and  $D=q+r+{\tilde d}+2=d+{\tilde d}+2$.

The equations of motion (\ref{111}) for a part of components of the
 antisymmetric field are identically satisfied and for the other part 
  they are reduced to a simple equation:

\begin{equation}
\partial _{\alpha}(e^{-\alpha \phi -2qA +C}\partial _{\alpha}C)=0.
                  \label{120}
\end{equation}

For $\alpha$-components of the antisymmetric field we 
 also have the Bianchi identity:

\begin{equation}
\partial_{\alpha}(e^{\alpha\phi + 2Bq + \chi }\partial_{\alpha}\chi)=0.=
  \label{121}
\end{equation}

The equation of motion for the dilaton has the form

$$\partial _{\alpha}(e^{qA +{\tilde d} B + Fr}\partial _{\alpha}\phi )
-\frac{\alpha h^{2}}{2}e^{-\alpha \phi -qA + qB +rF
+2C}(\partial _{\alpha} C)^{2}
$$

\begin{equation}
+\frac{\beta k ^{2}}{2}e^{\beta \phi +2{\tilde d}B+2\chi}
(\partial _{\alpha}\chi)^{2} =0.
  \label{122}
\end{equation}

We have to solve the system of equations (\ref{117})-(\ref{122})
for unknown functions$~~~~$
$ A,B,F,C,\chi,\phi$. We shall express $ A,B,F$ and  $\phi $
in terms of two functions $C$ and $\chi$. 
In order to get rid of  exponents in (\ref{117})-(\ref{122}) 
we impose the following relations:
\begin{equation}
              qA + rF +{\tilde d}B =0,
 \label{123}
\end{equation}

\begin{equation}
2\chi +2{\tilde d}B +\beta \phi  =0,
                    \label{124}
\end{equation}
\begin{equation}
 2C - 2qA -\alpha\phi = 0.
                       \label{125}
\end{equation}
Under these conditions equations (\ref{120}),(\ref{121}) and (\ref{122})
will have the following forms, respectively,

\begin{equation}
\partial _{\alpha}(e^{-C}\partial _{\alpha}C)=0,~~~~~
\partial _{\alpha}(e^{-\chi}\partial _{\alpha}\chi)=0,
                                   \label{126}
\end{equation}
\begin{equation}
\Delta \phi +\frac{\beta k ^{2}}{2}(\partial _{\alpha}  \chi  )^{2}- 
\frac{\alpha  h^{2}}{2}(\partial _{\alpha} C )^{2}=0.
                  \label{127}
\end{equation}

One rewrites (\ref{126}) as
\begin{equation}
 \Delta C =(\partial C)^{2},~~~~~
 \Delta \chi =(\partial \chi)^{2}.                  \label{128}
\end{equation}
Therefore (\ref{127}) will have the form
\begin{equation}
\Delta \phi +\frac{\beta k^{2}}{2}\Delta \chi - 
\frac{\alpha  h^{2}}{2}\Delta C =0.
                      \label{129}
\end{equation}

>From (\ref{129}) it is natural to set
\begin{equation}
\phi =\phi_{1}C + \phi_{2}\chi, 
 \label{130}
\end{equation}
where 
\begin{equation}
 \phi_{1}=\frac{\alpha h^{2}}{2},~~~
 \phi_{2}=-\frac{\beta k^{2}}{2}.
 \label{131}
\end{equation}

>From equations (\ref{123}), (\ref{124}) and (\ref{125})  it follows that 
$A$, $B$ and $F$ can 
be presented as 
linear combinations of functions $C$ and $\chi$:

\begin{equation}
 A=a_{1}C + a_{2}\chi  ,
 \label{132}
\end{equation}
\begin{equation}
 B=b_{1}C + b_{2}\chi  ,
 \label{133}
\end{equation}
\begin{equation}
 F=f_{1}C + f_{2}\chi  ,
 \label{134}
\end{equation}
where

\begin{equation}
 a_{1}=\frac{4-\alpha^{2}h^{2}}{4q}, ~~~
 a_{2} = \frac{\alpha\beta k^{2}}{4q},
 \label{135}
\end{equation}

\begin{equation}
 b_{1} = -\frac{\alpha\beta h^{2}}{4{\tilde d}},~~~
 b_{2} =\frac{\beta ^{2}k^{2}-4}{4{\tilde d}},
  \label{136}
\end{equation}

\begin{equation}
 f_{1} = \frac{\alpha^{2}h^{2} +\alpha \beta h^{2}-4}{4r},~~~
 f_{2} = \frac{4-\alpha \beta k^{2}-\beta^{2} k^{2}}{4r}.
 \label{137}
\end{equation}

Let us substitute expressions (\ref{130}),(\ref{132})-(\ref{134}) for
$\phi,A,B,F$ into (\ref{117})-(\ref{119}). We get relations containing 
bilinear forms over derivatives on $ C$ and $\chi$ .
 We assume that the
coefficients in front of these  bilinear
forms vanish. Then we get the system of twelve quartic equations 
which is presented and solved in Appendix.
The system has a solution only if
 $\alpha$ and $\beta$  satisfy the relation
\begin{equation}
\alpha \beta=\frac{2q{\tilde d}}{q+r+{\tilde d}}
 \label{81m}
\end{equation}
In this case $h$ and $k$ are given by the formulae
\begin{equation}
h=\pm\sqrt{\frac{4(q+r+{\tilde d})}
{\alpha ^{2}(q+r+{\tilde d})+2q({\tilde d}+r)}},
 \label{819}
\end{equation}
\begin{equation}
k=\pm\frac{2\alpha (q+r+{\tilde d})}{\sqrt{{\tilde d}
[2\alpha ^{2}(q+r)(q+r+{\tilde d})+4q^{2}{\tilde d}]}}
 \label{195}
\end{equation}
To summarize, the action (\ref{11}) has the solution of the form
(\ref{12}) expressed in terms
of two harmonic functions $H_{1}$ and $H_{2}$ 
if the parameters in the action are  related by (\ref{193})
and the parameters in the Ansatz $h$ and $k$ are given
by (\ref{819}),(\ref{195}).

\section {Discussion and Conclusion}
Let us discuss different particular cases of the solution (\ref{12}).
There is the relation (\ref{15}) between parameters $\alpha$
and $\beta$
in the action (\ref{011}). As a result the action
corresponds to the bosonic part of a supergravity theory
only in some dimensions. 

If $D=4$ and $ q=d=1$ then
one can take $\alpha=\beta=1$ and the action
corresponds to the $SO(4)$ version of $N=4$ supergravity.
The solution  (\ref{12}) takes the form
\begin{equation}
 ds^{2}=-H_{1}^{-1}H_{2}^{-1}dt^{2}+H_{1}H_{2}dx^{\alpha}dx^{\alpha}
 \label{195a}
\end{equation}
This supersymmetric solution has been obtained in \cite{kallosh}.

If $\alpha=\beta$ and $q=\tilde d$ then one has the solution
 \begin{equation}
 ds^{2}=H_{1}^{\frac{2}{D-2}}H_{2}^{\frac{2(D-q-2)}{q(D-2)}}
 [(H_{1}H_{2})^{-\frac{2}{q}} \eta_{\mu \nu} dy^{\mu} dy^{\nu}+
 H_{2}^{-\frac{2}{q}}dz^{m}dz^{m}+
 dx^{\alpha}dx^{\alpha}], 
  \label{713}
 \end{equation}
This solution was obtained in \cite{AV}. It contains
as a particular case for $d=10,~q=2$ the known solution
\cite{TS,CM,CH}
$$  ds^{2}=H_{1}^{-\frac{3}{4}}H_{2}^{-\frac{1}{4}}
  (-dy_{0}^{2}+dy_{1}^{2})
$$ 
\begin{equation}
 + H_{1}^{\frac{1}{4}}H_{2}^{-\frac{1}{4}}
  (dz_{2}^{2}+dz_{3}^{2}+dz_{4}^{2}+dz_{5}^{2})+
  H_{1}^{\frac{1}{4}}H_{2}^{\frac{3}{4}}
  (dx_{6}^{2}+dx_{7}^{2}+dx_{8}^{2}+dx_{9}^{2}).
  \label{66m}
 \end{equation}
This solution has been used in the D-brane derivation
of the black hole entropy \cite{SV,CM}. Note however
that the solution (\ref{66m}) corresponds to the action
(\ref{011})  with the $3$-form $F_{3}$ and the $7$-form
$G_{7}$.

To conclude, a rather general  three-block solution
of the action (\ref{011}) has been constructed. It contains
as particular cases many known solutions. However the solution
is not  general enough to include some known multi-block
solutions.  Further work is needed  to understand better
the structure hidden behind the multi-block p-brane solutions.

\section*{Acknowlegments}
This work is supported by an operating grant from 
the Natural Sciences and Engineering Research Council of Canada.
I.A. and I.V. thank the Department of Physics for the kind
hospitality during their stay at Simon Fraser University.
I.V. is partially supported by the RFFI grant 9600312.
 \section {Appendix}
 We obtain the following system of  algebraic equations

$$-a_{1}+b_{1}+\frac{q(q-1)}{2}a_{1}^{2}+\frac{r(r+1)}{2}f_{1}^{2}+
\frac{{\tilde d}({\tilde d}+1)}{2}b_{1}^{2}$$

\begin{equation}
 +{\tilde d}(q-1)a_{1}b_{1}+r(q-1)a_{1}f_{1}+r{\tilde d}f_{1}b_{1}+
 \frac{\phi_{1}^{2}}{4}+
 \frac{h^{2}}{4}=0;
 \label{178}
\end{equation}

$$-a_{2}+b_{2}+\frac{q(q-1)}{2}a_{2}^{2} + 
\frac{r(r+1)}{2}f_{2}^{2} + \frac{{\tilde d}({\tilde d}+1)}{2}b_{2}^{2}$$

\begin{equation}
 +{\tilde d}(q-1)a_{2}b_{2}+r(q-1)a_{2}f_{2}+r{\tilde d}b_{2}f_{2} 
 +\frac{\phi_{2}^{2}}{4}+\frac{k^{2}}{4}=0;
 \label{179}
\end{equation}

$$q(q-1)a_{1}a_{2} +r(r+1)f_{1}f_{2}+{\tilde d}({\tilde d}+1)b_{1}b_{2}$$

\begin{equation}
 +{\tilde d}(q-1)(a_{1}b_{2}+a_{2}b_{1})+r(q-1)(a_{2}f_{1}+a_{1}f_{2})+
 r{\tilde d}(f_{1}b_{2}
 +f_{2}b_{1})+\frac{\phi_{1}\phi_{2}}{2}=0;
 \label{180}
\end{equation}

$$-f_{1}+b_{1}+\frac{q(q+1)}{2}a_{1}^{2}+\frac{r(r-1)}{2}f_{1}^{2}+
\frac{{\tilde d}({\tilde d}+1)}{2}b_{1}^{2}$$

\begin{equation}
 +q {\tilde d} a_{1}b_{1}+q(r-1)a_{1}f_{1}+
 {\tilde d}(r-1)f_{1}b_{1}+\frac{\phi_{1}^{2}}{4}
 -\frac{h^{2}}{4}=0;
 \label{181}
\end{equation}

$$-f_{2}+b_{2}+\frac{q(q+1)}{2}a_{2}^{2} + 
\frac{r(r-1)}{2}f_{2}^{2} + \frac{{\tilde d}({\tilde d}+1)}{2}b_{2}^{2}+$$

\begin{equation}
 +q {\tilde d} a_{2}b_{2}+q(r-1)a_{2}f_{2}+{\tilde d}(r-1)b_{2}f_{2} 
 +\frac{\phi_{2}^{2}}{4}+\frac{k^{2}}{4}=0;
 \label{182}
\end{equation}

$$q(q+1)a_{1}a_{2} +r(r-1)f_{1}f_{2}+{\tilde d}({\tilde d}+1)b_{1}b_{2}$$

\begin{equation}
 +q{\tilde d}(a_{1}b_{2}+a_{2}b_{1})+q(r-1)(a_{2}f_{1}+a_{1}f_{2})
 +{\tilde d}(r-1)(f_{1}b_{2}
 +f_{2}b_{1})+\frac{\phi_{1}\phi_{2}}{2}=0;
 \label{183}
\end{equation}

$$-qa_{1}^{2}-rf_{1}^{2}+
{\tilde d} b_{1}^{2}+$$

\begin{equation}
 +2qa_{1}b_{1}+2rf_{1}b_{1}-\frac{\phi_{1}^{2}}{2}+
 \frac{h^{2}}{2}=0;
 \label{184}
\end{equation}

$$-qa_{2}^{2} -rf_{2}^{2} + {\tilde d}b_{2}^{2}$$

\begin{equation}
 +2qa_{2}b_{2}+2rb_{2}f_{2} 
 -\frac{\phi_{2}^{2}}{2}+\frac{k^{2}}{2}=0;
 \label{185}
\end{equation}

$$-qa_{1}a_{2} -rf_{1}f_{2}+{\tilde d}b_{1}b_{2}$$

\begin{equation}
 +q(a_{1}b_{2}+a_{2}b_{1})+r(f_{1}b_{2}
 +f_{2}b_{1})-\frac{\phi_{1}\phi_{2}}{2}=0;
 \label{186}
\end{equation}

$$\frac{q(q+1)}{2}a_{1}^{2}+\frac{r(r+1)}{2}f_{1}^{2}+
\frac{{\tilde d}({\tilde d}-1)}{2}b_{1}^{2}$$

\begin{equation}
 +q({\tilde d}-1)a_{1}b_{1}+r({\tilde d}-1)b_{1}f_{1}+
 rqf_{1}a_{1}+\frac{\phi_{1}^{2}}{4}-
 \frac{h^{2}}{4}=0;
 \label{187}
\end{equation}

$$\frac{q(q+1)}{2}a_{2}^{2} + 
\frac{r(r+1)}{2}f_{2}^{2} + \frac{{\tilde d}({\tilde d}-1)}{2}b_{2}^{2}$$

\begin{equation}
 +q({\tilde d}-1)a_{2}b_{2}+r({\tilde d}-1)a_{2}f_{2}+rqb_{2}f_{2} 
 +\frac{\phi_{2}^{2}}{4}-\frac{k^{2}}{4}=0;
 \label{188}
\end{equation}

$$q(q+1)a_{1}a_{2} +r(r+1)f_{1}f_{2}+{\tilde d}({\tilde d}-1)b_{1}b_{2}+
q({\tilde d}-1)(a_{1}b_{2}+a_{2}b_{1})
$$

\begin{equation}
+r({\tilde d}-1)(b_{2}f_{1}+b_{1}f_{2})+rq(f_{1}a_{2}
 +f_{2}a_{1})+\frac{\phi_{1}\phi_{2}}{2}=0.
 \label{188a}
\end{equation}

Now let us discuss the system of eq. (\ref{178})-(\ref{188a}).
The action  depends on the parameters $D,\alpha, \beta ,q,\tilde{d}$.
We have used the parameter $r$ instead of $D$ which is defined from
$$q+r+\tilde{d} +2=D$$
Therefore our action (\ref{11}) depends on five parameters
$r,\alpha, \beta ,q,\tilde{d}$
We have also two parameters $h$ and $k$ in our Ansatz 
(\ref{115}),(\ref{116}).
We substitute expressions (\ref{130}),(\ref{131}),
(\ref{135})-(\ref{137})
into  (\ref{178})-(\ref{188a}). Then we get
the system of twelve 
quartic equations for
seven unknown variables $r,\alpha, \beta ,q,\tilde{d},h$ and $k$. 
Using  Maple V we found the solution of the system 
of equations (\ref{178})-(\ref{188a}). The solution has the form
\begin{equation}
k=\frac{\sqrt{2q(r+{\tilde d})h^{2}-4(q+r+{\tilde d})}}
{\sqrt{{\tilde d}(qrh^{2}-2(q+r))}},
 \label{190}
\end{equation}
\begin{equation}
\alpha=\frac{\sqrt{4(q+r+{\tilde d})-2q(r+{\tilde d})h^{2}}}
{h\sqrt{q+r+{\tilde d}}}
 \label{191}
\end{equation}
\begin{equation}
\beta=\frac{2q{\tilde d}h}{\sqrt{(q+r+{\tilde d})
[4(q+r+{\tilde d})-2q(r+{\tilde d})h^{2}]}}
 \label{192}
\end{equation}
where $r,q,\tilde{d},h$ are arbitrary. Let us rewrite
the solution in terms of parameters in the action.
>From equations (\ref{191}) and (\ref{192}) follows that 
$h$ can be found from equations (\ref{178})-(\ref{188a}) only
if $\alpha$ and $\beta$  are subjected to the relation
\begin{equation}
\alpha \beta=\frac{2q{\tilde d}}{q+r+{\tilde d}}
 \label{193}
\end{equation}
In this case $h^{2}$ and $k^{2}$ are given by the formulae
\begin{equation}
h^{2}=\frac{4(q+r+{\tilde d})}
{\alpha ^{2}(q+r+{\tilde d})+2q({\tilde d}+r)},
 \label{}
\end{equation}
\begin{equation}
k^{2}=\frac{2\alpha^{2} (q+r+{\tilde d})^{2}}
{{\tilde d}[\alpha ^{2}(q+r)(q+r+{\tilde d})+2q^{2}{\tilde d}]}
 \label{}
\end{equation}

\end{document}